\documentclass[prb,twocolumn,showpacs,noshowkeys,
preprintnumbers,amsmath,amssymb]{revtex4}
\usepackage{graphicx}% Include figure files
\usepackage{dcolumn}% Align table columns on decimal point
\usepackage{bm}% bold math
\usepackage{hyperref}%
\begin{document}

\preprint{}

\title{Critical generalized inverse participation ratio distributions}
\author{E. Cuevas}
\email{ecr@um.es}
\homepage{http://bohr.fcu.um.es/miembros/ecr/}
\affiliation{Departamento de F{\'\i}sica, Universidad de Murcia,
E-30071 Murcia, Spain}

\date{\today}

\begin{abstract}
The system size dependence of the fluctuations in generalized inverse
participation ratios (IPR's) $I_{\alpha}(q)$ at criticality is investigated
numerically. We focus on a three-dimensional (3D) system with unitary
symmetry, a 2D system with symplectic symmetry and a 1D system with
orthogonal symmetry. The variances of the IPR logarithms are found to be
scale-invariant at the macroscopic limit. The finite size corrections
to the variances decay algebraically with nontrivial exponents, which
depend on the Hamiltonian symmetry and the dimensionality. The large-$q$
dependence of the asymptotic values of the variances behaves as $q^2$
according to theoretical estimates. These results ensure the self-averaging
of the corresponding generalized dimensions.
\end{abstract}


%%%\keywords{Suggested keywords}

\maketitle

The statistical properties of wave functions have been intensively studied
during recent years, helping in our understanding of phenomena in different
areas of physics, ranging from nuclear \cite{ZB96} and atomic \cite{BU96}
to microwave \cite{PS00} and mesoscopic physics. \cite{AK91,MK95,FM95,AL97,
Mi00,Ch96} Of considerable interest is what happens at the critical point
of a metal-insulator transition (MIT), in which eigenstates presenting
strong amplitude fluctuations were demonstrated to be multifractal objects.

The Anderson MIT essentially depends on the dimensionality and
symmetries of the system. The field theoretical formulation \cite{Ef83}
of the MIT shows that the critical behavior can be described within a
framework of three universality classes: orthogonal ($\beta=1$), unitary
($\beta=2$), and symplectic ($\beta=4$). The symmetry parameter $\beta$
is the number of independent real components that characterizes a matrix
element of the corresponding Hamiltonian. The two major symmetries in
field theory are time-reversal symmetry and spin-rotation symmetry.
The system belongs to the orthogonal class if it has both symmetries, to
the unitary class if time-reversal symmetry is broken, and to the symplectic
class if the system has time-reversal symmetry but spin-rotation is broken.
The relevant terms in the hamiltonian are a coupling to an applied magnetic
field, which breaks time-reversal symmetry, and the spin-orbit interaction,
which breaks spin-rotation symmetry.

Castellani and Peliti \cite{CP86} considered the $d=2+\epsilon$
($\epsilon \ll 1$) expansion of Wegner's nonlinear $\sigma$ model
\cite{W80} for the generalized inverse participation ratios (IPR's)
of eigenstates,
\begin{equation}
I_{\alpha}(q)=\int_\Omega d^dr\;|\psi_\alpha({\mathbf r})|^{2q}\
\propto L^{-d_q(q-1)}\;,\label{ipr}
\end{equation}
and concluded that the wave functions $\psi_\alpha({\mathbf r})$ at the
transition point of a MIT show multifractal behavior characterized by an
infinite set of generalized fractal dimensions $d_q$. The index $\alpha$
labels the different eigenfunctions and $\Omega$ denotes a $d$-dimensional
region with linear dimension $L$. Equation (\ref{ipr}) is valid for individual
states and for their ensemble average since the spectrum of multifractal
dimensions has universal features for states in the vicinity of the MIT
\cite{Ja94}.

As regards the IPR fluctuations, we shall first review the existing analytical
results. When these fluctuations were studied for two-dimensional systems in the
framework of the supersymmetry method, \cite{FM95,PA98,Mi00} it was found that
the distribution function of $I_{\alpha}(q)$, normalized to its typical value,
is scale-invariant at criticality. \cite{PA98} Although this case does not
present a \textit{true} Anderson transition, the above result motivated the
conjecture that in general the normalized distribution function of $I_{\alpha}(q)$
is universal, i.e., size independent for $L\to\infty$. Accordingly, it is assumed
that the distribution function of $\ln I_{\alpha}(q)$ is a universal curve that,
while horizontally shifted by changes in $L$, keeps the same form.
Within the same approach, Mirlin
and Evers \cite{EM00,ME00} study the fractal properties of the power-law random
banded matrix (PRBM) ensemble at criticality. This model is characterized by a
parameter, $b$ [see Eq. (\ref{h1dor}) below]. In the two limiting cases, $b \gg 1$
and $b \ll 1$, for which analytical treatment is feasible, they found the
distribution function of the IPR to be scale-independent at the macroscopic
limit. Particularly, they analyzed the behavior of the variance $\sigma^2_{z_q}$
of the distribution function of $I_{\alpha}(q)$, normalized to its typical value,
in the limit $b \ll 1$. For values of $q$ larger than some critical value,
$q_{\rm c} \approx 2.41$, where the IPR distribution is dominated by its slowly
decaying power law tail, the same authors obtained for $L \to \infty$
\begin{equation}
\sigma^2_{z_q} (\infty)=\left( q/q_{\rm c} \right)^2\;.
\label{varq}
\end{equation}
What we are trying to point out is that a general analytical description of
the statistical properties of the generalized IPR at the critical point of
realistic MIT's (conventional Anderson transition, quantum Hall transition,
transition in $d=2$ for electrons with strong spin-orbit coupling, etc.) is
still lacking. This is why we addressed the problem using numerical
calculations. To our knowledge, no such calculations have been previously
reported.

Our aim was to find the system size dependence of the generalized
IPR distributions in order to check whether they converge at large
$L$ to the conjectured scale invariant distributions or not. A second
motivation for this work was to clarify whether the corresponding
generalized dimensions are well-defined in the macroscopic limit or
presents scale-invariant distributions.

Using the exact eigenstates from numerical diagonalizations, we
obtained the size dependence of the distribution function
${\cal P}_{\beta}(z_q=\ln I_{\alpha}(q))$ at the critical point of a
three-dimensional (3D) system with unitary symmetry, a two-dimensional
(2D) system with symplectic symmetry and a one-dimensional (1D) system
with orthogonal symmetry. All these systems possess a mobility edge and
are modeled by the corresponding tight-binding Hamiltonian. We found
that as $L$ increases ${\cal P}_{\beta}$ becomes wider and the height of
its peak smaller with a tendency towards saturation. Therefore, their
variance increases as $L$ increases and will eventually saturate to a
constant finite value for $L \to \infty$. In order to extrapolate to
macroscopic systems, we propose a finite size correction to the variance
$\sigma^2_{z_q}$ of $z_q$ according to
\begin{equation}
\sigma^2_{z_q}(L)=\sigma^2_{z_q}(\infty)-a_q L^{-y_q}\;,\label{s2l}
\end{equation}
with $\sigma^2_{z_q}(\infty)$, $a_q$ and  $y_q > 0$ being three adjustable
parameters. Practically for all triads $\{q,\beta,d \}$ we found that the
exponent $y_q$ was always close to the generalized dimension $d_q$ divided
by $2\beta d$. Thus, for the exponent $y_q$ we propose the following equation,
\begin{equation}
y_q=\frac{d_q}{2\beta d}\;,\label{yq}
\end{equation}
and keep only two free parameters. This relation constitutes the main result
of the present publication. As we will see below, our numerical data strongly
support this conjecture. Equation (\ref{yq}) is a generalization of the exponent
$\gamma=d_2/2d$ found by our group \cite{CO02} for $I_{\alpha}(2)$ in
disordered systems with orthogonal symmetry.

Some technical details follow: the critical eigenfunctions (and
eigenvalues) of the Hamiltonian matrices, Eqs. (\ref{h3dun}), (\ref{h2dsp}),
and (\ref{h1dor}), are obtained by numerical diagonalization. In 3D and 2D
systems we use techniques for large sparse matrices, \cite{CW85} while for
the 1D case standard diagonalization subroutines are used, since we must
deal with full matrices. We consider a small energy window $(-\eta,\eta)$
around the center of the spectral band, where $\eta=1$ ($\eta=0.4$) for 3D
and 2D (1D) systems. Reducing the width of the above windows does not alter
the results. For an accurate extrapolation to the thermodynamic limit, the system
size changes by one order of magnitude for 3D and 2D systems and by two orders
in the case of 1D. The number of random realizations is such that the number
of critical states included for each $L$ is roughly $1.5\times 10^5$, while,
in order to reduce edge effects, we impose periodic boundary conditions in all
the cases considered. We have checked that the normalized nearest level variances
\cite{Cu99} are indeed scale-invariant at each critical point studied. Below we
present our numerical results and compare them to the conjecture (\ref{yq}).

\textit{3D unitary symmetry}. First we focus on the breaking
of time-reversal symmetry by a constant applied magnetic field.
We consider the standard Anderson model with diagonal disorder
and Peierls phase factors in the hopping matrix elements 
\begin{equation}
{\cal H} = \sum_{n}\epsilon_n c_{n}^{\dagger}c_{n}+ \sum_{n,m} V_{nm}
c_{n}^{\dagger}c_{m} 
\label{h3dun}\;,
\end{equation}
where $c_{n}^{\dagger}$ ($c_{n}$) are the creation (annihilation)
operators of an electron at a site $n$ of a simple-cubic lattice,
and the summation in the second term is over nearest neighbors.
The diagonal energies are randomly distributed with constant probability
in the interval $-W/2<\epsilon_n <W/2$, where $W$ parametrizes the strength
of random potential. Energies are measured in units of the modulus of the
hopping matrix elements, $V_{nm}$, and lengths in units of the lattice constant.
The hopping matrix elements, $V_{nm}=e^{\mp 2\pi i \alpha z}$ for neighbors
in the $y$ direction and $V_{nm}=1$ in any other direction, describe a system
with a homogeneous magnetic field, $B$, in the $x$ direction, while the Peierls
phase $\alpha =eB/hc$ is the number of flux quanta $\Phi_{0}=hc/e$ threading a
lattice cell. Here, the Landau gauge with the vector potential
$\mathbf{A}=(0,-Bz,0)$ is chosen. The system size varies between 4 and 30,
and for the critical disorder we take the value $W_{\rm c}=18.35$ reported
in Ref. %\cite{DB98}%
20 for $\alpha=0.25$.

\begin{figure}
\begin{center}
\includegraphics[width=8.0cm]{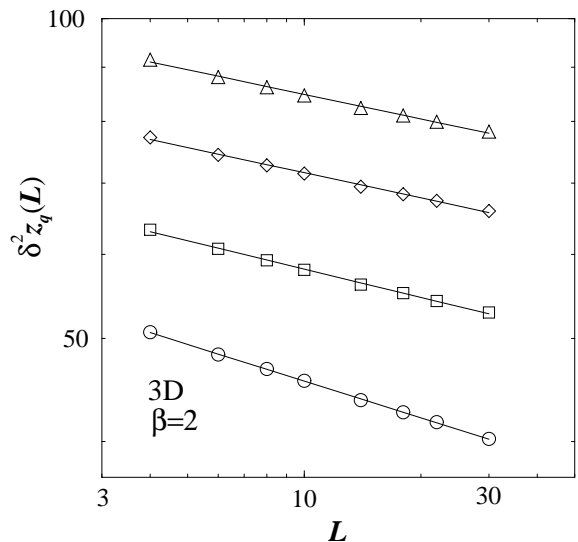}
\caption{\label{fig1} Finite size correction $\delta^2 z_q (L)$ as a
function of system size $L$ for the 3D disordered system with unitary
symmetry and several $q$ values: $q=2$ (circles), $q=4$ (squares), $q=6$
(diamonds), and $q=8$ (triangles). The straight lines are linear fits to
Eq. (\ref{s2l}).}
\end{center}
\end{figure}

In Fig. 1 we represent on a log-log scale the finite size correction
$\delta^2 z_q (L) \equiv \sigma^2_{z_q}(\infty)-\sigma^2_{z_q}(L)$
as a function of $L$ for the 3D disordered system with unitary symmetry and
different values of $q$: $2$ (circles), $4$ (squares), $6$ (diamonds), and $8$
(triangles). The good agreement with Eqs. (\ref{s2l}) and (\ref{yq}) is evident.
The fitted values of the free parameters are $a_2=2.01$, $a_4=19.71$,
$a_6=52.88$, $a_8=100.7$, while the corresponding $\sigma^2_{z_q}(\infty)$ are
summarized in Fig. 4 (squares). The slopes of the straight lines have not been
fitted and correspond to Eq. (\ref{yq}), with $\beta=2$, $d=3$, $d_2=1.37$,
$d_4=1.07$, $d_6=0.98$, and $d_8=0.94$. These generalized dimensions were
estimated from the slope, $d_q(1-q)$, of $\langle z_q \rangle$ versus $\ln L$.
In order to produce a diagram of reasonable size the curves for $q=2$, 4, and 6
have been shifted vertically.
The small difference in the slopes for $q \ge 4$ correspond to the slight
differences in the values of $d_q$. The value $\sigma_{z_2}(\infty)=1.40$ found
is in good agreement with the conjecture $\sigma_{z_2}(\infty)\sim 1$. \cite{FM95}

\textit{2D symplectic symmetry}. In order to calculate the critical eigenstates
(two-component spinors) Ando's model Hamiltonian \cite{An89} was used,
\begin{equation}
{\cal H} = \sum_{n,\sigma}\epsilon_n c_{n,\sigma}^{\dagger}c_{n,\sigma}+
\sum_{n,\sigma,m,\sigma'} V_{n,\sigma;m,\sigma'}
c_{n,\sigma}^{\dagger}c_{m,\sigma'} 
\label{h2dsp}\;,
\end{equation}
where $c_{n,\sigma}^{\dagger}$ ($c_{n,\sigma}$) are the creation (annihilation)
operators of an electron at a site $n$ of a square lattice  with the spin
component $\sigma$ and $m$ denotes the sites adjacent to site $n$. The on-site
energy $\epsilon_n$ is randomly distributed around zero according to a box
distribution with a width $W$, which specifies the degree of disorder. The index
$\sigma$ takes on values of $+1$ and $-1$, and denotes spin-up or spin-down. The
transfer matrices $V_{n,\sigma;m,\sigma'}=V_x$ or $V_y$ depend on the direction
of the nearest neighbor site
\begin{equation} \nonumber
V_{x}=\left( \begin{array}{cc}
V_{1} & V_{2} \\ -V_{2} & V_{1} \end{array} \right), \quad
V_{y}=\left( \begin{array}{cc}
V_{1} & -iV_{2} \\ -iV_{2} & V_{1} \end{array} \right),
\label{vxvy}
\end{equation}
and the strength of the spin-orbit coupling is given by the parameter 
$S=V_2/V$, with $V=(V_1^2+V_2^2)^{1/2}$ taken to be the unit of energy.
We chose $S=1/2$ and, for the critical disorder, $W_{\rm c}=5.98$, which
was previously found in Ref. 22. %\cite{SZ97}%
The system size ranges from
$L=10$ to $L=120$.

\begin{figure}
\begin{center}
\includegraphics[width=8.0cm]{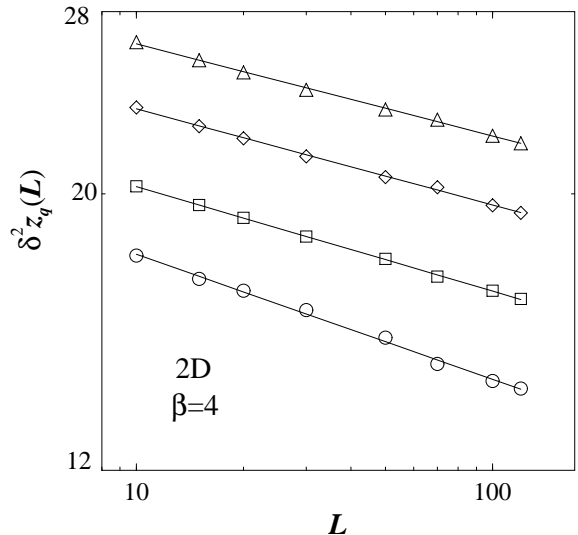}
\caption{\label{fig2} As for Fig. 1, for the 2D disordered system with
symplectic symmetry.}
\end{center}
\end{figure}

Figure 2 shows the same size corrections as Fig. 1 for the 2D disordered
system with symplectic symmetry. In this case the fitted values of the
free parameters are $a_2=0.17$, $a_4=5.10$, $a_6=15.77$, and $a_8=31.27$ and
the values of $\sigma^2_{z_q}(\infty)$ are shown in Fig. 4 (circles).
As in the 3D case, the slopes of the straight lines correspond to
Eq. (\ref{yq}), with $\beta=4$, $d=2$, $d_2=1.60$, $d_4=1.34$, $d_6=1.23$,
and $d_8=1.18$. These dimensions were also estimated in a similar way.
Again, the value found, $\sigma_{z_2}(\infty)=0.46$, is in agreement
with the conjecture $\sigma_{z_2}(\infty)\sim 1$. \cite{FM95}

\textit{1D orthogonal symmetry}.
In this case the PRBM model \cite{EM00,ME00} is used. The Hamiltonian,
which describes a disordered 1D sample with random long-range hopping,
is represented by real symmetric matrices, whose entries are randomly
drawn from a normal distribution with zero mean
$\left\langle {\cal H}_{nm} \right\rangle =0$, and a variance depending
on the distance between lattice sites
\begin{equation}
\left\langle |{\cal H}_{nm}|^2\right\rangle =\frac{1}{1+(|n-m|/b)^2} 
\label{h1dor}\;.
\end{equation}
The model describes a whole family of critical theories parametrized
by $b$, which determines the critical dimensionless conductance, in the
same way as the dimensionality labels the different Anderson transitions.
In the intermediate regime $b \sim 1$, where there is no analytical
solution, we obtained two different diverging exponents for the
correlation and localization lengths at the critical region. \cite{CG01}
The system size ranges between $L=100$ and $L=10\,000$, and for the
calculations we choose $b=0.2$.

\begin{figure}
\begin{center}
\includegraphics[width=8.0cm]{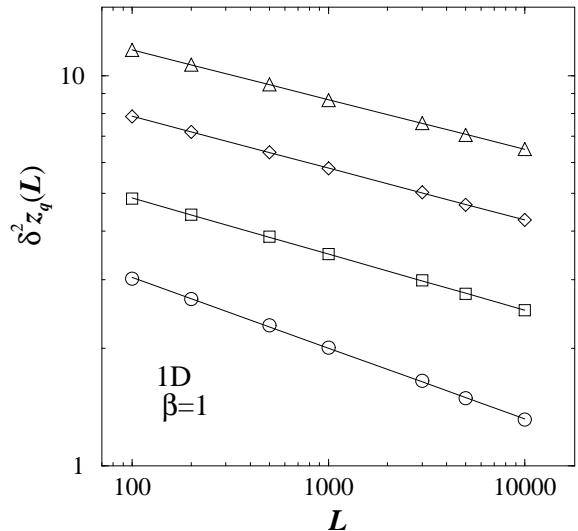}
\caption{\label{fig3} As for Fig. 1, for the 1D disordered system with
orthogonal symmetry.}
\end{center}
\end{figure}

Figure 3 shows the same size corrections as Fig. 1 for the 1D disordered
system with orthogonal symmetry. The fitted values of the free parameters
are $a_2=0.47$, $a_4=4.28$, $a_6=11.16$, $a_8=20.95$, and
$\sigma^2_{z_q}(\infty)$ are shown in Fig. 4 (diamonds). As in the
two previous cases, the slopes of the straight lines correspond to
Eq. (\ref{yq}), with $\beta=1$, $d=1$, $d_2=0.36$, $d_4=0.29$,
$d_6=0.27$, and $d_8=0.26$. The last four dimensions were estimated in
a way similar to those for Fig. 1. As in the 3D and 2D cases the value
$\sigma_{z_2}(\infty)=0.67$ found is in agreement with the conjecture
$\sigma_{z_2}(\infty) \sim 1$. \cite{FM95}

We checked that Eqs. (\ref{s2l}) and (\ref{yq}) adequately describe
the IPR fluctuations in the standard 3D Anderson transition (orthogonal
symmetry) and in the critical point of the 1D system with long-range
hopping, Eq. (\ref{h1dor}), when unitary and symplectic symmetries are
included. We are confident that further analytical development in IPR
statistics at criticality will confirm our conjectured result.

The statistical properties of wave functions are intimately
related to the universal conductance fluctuations. As was shown
in Ref. 6 %\cite{FM95}%
, the relative variance of $I_\alpha(2)$ is
of the order of $1/g^2$, $g$ being the dimensionless conductance
(in units of $e^2/h$). The extrapolation of this relation to the
transition point, in which $g\sim 1$, is the most direct connection
between our results and experiments. In this way, our results could
be checked in experiments with random systems (semiconductor
heterostructures, metal oxide semiconductor MOS inversion layers,
vapor-deposited films, doped and amorphous semiconductor, etc.) using
samples with varying  size.

\begin{figure}
\begin{center}
\includegraphics[width=8.0cm]{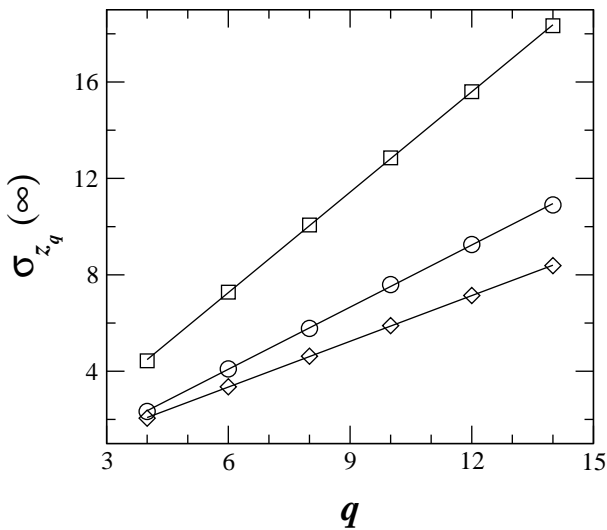}
\caption{\label{fig4} Asymptotic values of the standard deviation
$\sigma_{z_q}(\infty)$ as a function of $q$ for the 1D system with
orthogonal symmetry (diamonds), 2D symplectic (circles), and 3D unitary
(squares).}
\end{center}
\end{figure}

Finally, we focus on the $q$ dependence of the asymptotic values
$\sigma^2_{z_q}(\infty)$ for each universality class. In Fig. 4 we
represent the standard deviation $\sigma_{z_q}(\infty)$ versus $q$
for the three transitions studied: 1D orthogonal (diamonds), 2D
symplectic (circles), and 3D unitary (squares). The slopes of the
fitted straight lines are 0.63, 0.86, and 1.39, respectively. It is
interesting to note that the nonlinear $\sigma$ model estimates,
Eq. (\ref{varq}), of the 1D PRBM model is valid for \textit{all} three
transitions.  These results further support our conjecture, described
by Eq. (\ref{yq}). The different values of the slopes directly reflect
the universality class and the dimensionality of the system.
From the above slope for the 1D system we obtain $q_{\rm c}=1.59$,
which is less than the value of 2.41 predicted theoretically.
The origin of the deviation is that the latter value of $q_{\rm c}$
was obtained for the distribution function of $I_{\alpha}(q)$
normalized to its typical value, while the former $q_{\rm c}$
corresponds to the distribution function of $\ln I_{\alpha}(q)$.
Anyway, we have estimated the value of $q_{\rm c}$ from the
extrapolated long tails of the normalized $I_{\alpha}(q)$
distributions. The obtained value $q_{\rm c}=2.21$ is in close
agreement with the predicted one for these distributions.

If we now assume that the fluctuations of the generalized dimensions $d_q$
are due to fluctuations of the corresponding generalized IPR, \cite{Kr96}
from Eq. (\ref{ipr}) we obtain
$\sigma^2_{d_q}(L)=\sigma^2_{z_q} (L)/(q-1)^2\ln^2 L\;$
for the variance of $d_q$.
From this relation, and taking into account Eqs. (\ref{varq}), (\ref{s2l}),
and (\ref{yq}), we see that $\sigma_{z_q} (L)$ vanishes at a rate $1/\ln L$
when $L \to \infty$. Therefore, we conclude that all generalized dimensions
are self-averaged quantities, i.e., they tend to well-defined single values
at the macroscopic limit.

In summary, we have performed a detailed numerical analysis of the
statistical properties of the IPR distributions at the critical point
of three standard MIT's, finding the finite asymptotic values of the
variances of the distributions in each case. The corresponding finite
size corrections decay with system size as a power law with exponents
that depend on the system dimensionality and on the Hamiltonian symmetry.
For large values of $q$, the asymptotic
values $\sigma^2_{z_q}(\infty)\propto q^2$ for all three transitions is
in agreement with the analytical estimates of the 1D PRBM model. Therefore,
all generalized dimensions are well-defined at the macroscopic limit. 

%%%\begin{acknowledgments}
We would like to thank the Spanish DGESIC, Project No. 1FD97-1358
and No. BFM2000-1059, for financial support. 
%%%\end{acknowledgments}

\end{document}